%% file: main.tex
\documentclass[sigconf]{acmart}

\usepackage{booktabs} 
\usepackage{subcaption}
\usepackage{tabularx,color}
\usepackage{graphicx,eqnarray,amsmath,multirow,multicol,url,enumitem}
\usepackage[acronym,smallcaps,nowarn]{glossaries}

\newacronym{DGRec}{DGRec}{Dynamic Graph Recommendation}

\copyrightyear{2019}
\acmYear{2019}
\setcopyright{acmcopyright}
\acmConference[WSDM '19]{The Twelfth ACM International Conference on Web Search and Data Mining}{February 11--15, 2019}{Melbourne, VIC, Australia} \acmBooktitle{The Twelfth ACM International Conference on Web Search and Data Mining (WSDM '19), February 11--15, 2019, Melbourne, VIC, Australia} \acmPrice{15.00}
\acmDOI{10.1145/3289600.3290989}
\acmISBN{978-1-4503-5940-5/19/02}

\begin{document}

\title{Session-based Social Recommendation via Dynamic Graph Attention Networks}

\author{Weiping Song}
\affiliation{
  \institution{School of EECS, Peking University}}
\email{songweiping@pku.edu.cn}
\author{Zhiping Xiao}
\affiliation{
  \institution{School of EECS, UC Berkeley}}
\email{patricia.xiao@berkeley.edu}
\author{Yifan Wang}
\affiliation{
  \institution{School of EECS, Peking University}}
\email{yifanwang@pku.edu.cn}
\author{Laurent Charlin}
\affiliation{
  \institution{Mila \& HEC Montreal }}
\email{laurent.charlin@hec.ca}
\author{Ming Zhang}\authornote{Corresponding authors.}
\affiliation{
  \institution{School of EECS, Peking University}}
\email{mzhang_cs@pku.edu.cn}
\author{Jian Tang}\authornotemark[1]
\affiliation{
  \institution{Mila \& HEC Montreal}}
\email{jian.tang@hec.ca}

\renewcommand{\shortauthors}{W. Song et al.}

\begin{abstract}
Online communities such as Facebook and Twitter are enormously popular and have become an essential part of the daily life of many of their users. Through these platforms, users can discover and create information that others will then consume. In that context, recommending relevant information to users becomes critical for viability. However, recommendation in online communities is a challenging problem: 1) users' interests are dynamic, and 2) users are influenced by their friends. Moreover, the influencers may be context-dependent. That is, different friends may be relied upon for different topics.
Modeling both signals is therefore essential for recommendations.

We propose a recommender system for online communities based on a dynamic-graph-attention neural network. We model dynamic user behaviors with a recurrent neural network, and context-dependent social influence with a graph-attention neural network, which dynamically infers the influencers based on users' current interests.
The whole model can be efficiently fit on large-scale data. Experimental results on several real-world data sets demonstrate the effectiveness of our proposed approach over several competitive baselines including state-of-the-art models. The source code and data are available at \url{https://github.com/DeepGraphLearning/RecommenderSystems}.

\end{abstract}



\maketitle

\input{intro}
\input{related}
\input{definition}

\input{model}
\input{experiment}
\input{conclusion}
\input{ack}

\bibliographystyle{ACM-Reference-Format}
\bibliography{reference}

\end{document}

%% file: intro.tex
\section{Introduction}
\label{sec::intro}

\noindent Online social communities are an essential part of today's online experience. 
Platforms such as Facebook, Twitter, and Douban enable users to create and share information as well as consume the information created by others. Recommender systems for these platforms are therefore critical to surface information of interest to users and to improve long-term user engagement. However, online communities come with extra challenges for recommender systems.

\begin{figure*} [!ptb]
\centering
\includegraphics[width=0.95\linewidth]{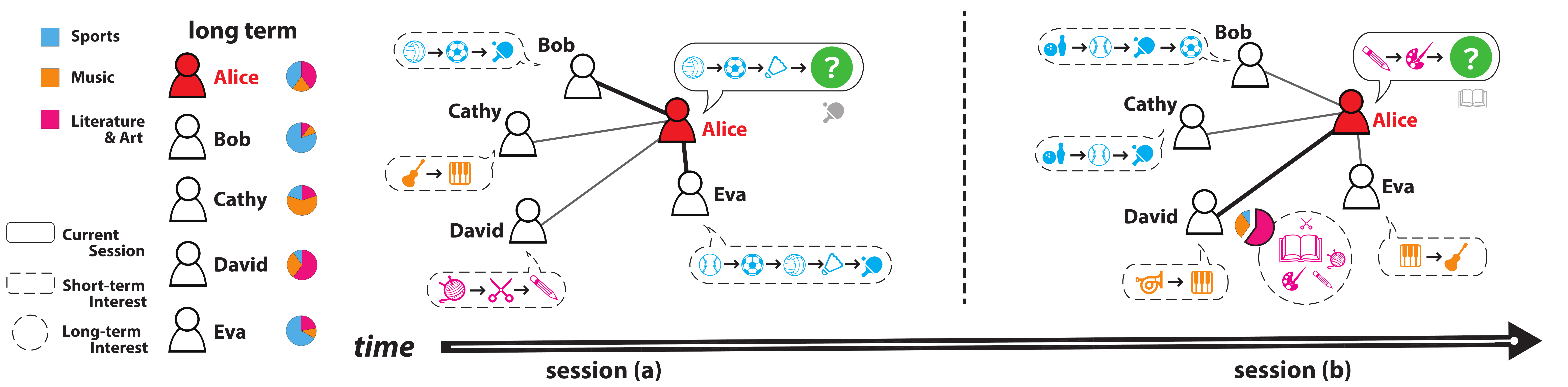}
\caption{An illustration of Alice's social influences in two sessions. Alice's interests might change across different sessions, while she may be influenced by her friends, by either their short-term or long-term preferences at different times.
}
\label{fig::motivation}
\end{figure*}

First, user interests are dynamic by nature. A user may be interested in sports items for a period of time and then search for new music groups. Second, since online communities often promote sharing information among friends, users are also likely to be influenced by their friends. For instance, a user looking for a movie may be influenced by what her friends have liked. 
Further, the set of influencers can be dynamic since they can be context-dependent. For instance, a user will trust a set of friends who like comedies when searching for funny films; while she could be influenced by another set of friends when searching for action movies.

\noindent \textbf{Motivating Example.} Figure~\ref{fig::motivation} presents the behavior of Alice's and her friends' in an online community. Behaviors are described by a sequence of actions (e.g., item clicks). To capture users' dynamic interests, their actions are segmented into sub-sequences denoted as \emph{sessions}. We are therefore interested in \emph{session-based recommendations}~\cite{schafer1999recommender}: within each session, we recommend the next item Alice should consume based on the items in the current session she has consumed so far.
Figure~\ref{fig::motivation} presents two sessions: session (a) and (b). 
In addition, the items consumed by Alice's friends are also available. We would like to utilize them for better recommendations. We are thus in a \emph{session-based social recommendation} setting.

In session (a), Alice browses sports items.
Two of her friends: Bob and Eva, are notorious sports fans (long-term interests), and they are browsing sports' items recently (short-term interests). Considering both facts, Alice may be influenced by the two and, e.g., decides to learn more about Ping Pong next. In session (b), Alice is interested in ``literature \& art'' items. The situation is different with session (a) since none of her friends have consumed such items recently. But David is generally interested in this topic (long-term interests). In this case, it would make sense for Alice to be influenced by David, and say, be recommended a book that David enjoyed.
These examples show how a user's current interests combining with the (short- and long-term) interests of different friends' provide session-based social recommendations. In this paper, we present a recommendation model based on both.

The current recommendation literature has modeled either users' dynamic interests or their social influences, but, as far as we know, has never combined both (like in the example above). A recent study~\cite{hidasi2016session} models session-level user behaviors using recurrent neural networks, ignoring social influences. Others studied merely social influences~\cite{ma2011recommender,zhao2014leveraging,chaney2015probabilistic}. For example, \citet{ma2011recommender} explores the social influence of friends' long-term preferences on recommendations. However, the influences from different users are static, not depicting the users' current interests.

We propose an approach to model both users' session-based interests as well as dynamic social influences. That is, which subset of a user's friends influence her (the influencers) according to her current session.
Our recommendation model is based on dynamic-graph-attention networks. Our approach first models user behaviors within a session using a recurrent neural network (RNN)~\cite{elman1990finding}. 
According to user's current interests---captured by the hidden representation of the RNN---we capture the influences of friends using the graph-attention network~\cite{Velickovic2018graph}. To provide session-level recommendations, we distinguish the model of friends' short-term preferences from that of the long-term ones.
The influence of each friend, given the user's current interests, is then determined automatically using an attention mechanism~\cite{bahdanau2015neural,xu2015show}.

We conduct extensive experiments on data sets collected from several online communities (Douban, Delicious, and Yelp). Our proposed approach outperforms the well-known competitive baselines by modeling both users' dynamic behaviors and dynamic social influences.

To summarize, we make the following contributions:
\begin{itemize}
\item We propose to study both dynamic user interests and context-dependent social influences for the recommendation in online communities.
\item We propose a novel recommendation approach based on dynamic-graph-attention networks for modeling both dynamic user interests and context-dependent social influences. The approach can effectively scale to large data sets. 
\item We conduct extensive experiments on real-world data sets. Experimental results demonstrate the effectiveness of our model over strong and state-of-the-art baselines.
\end{itemize}

\noindent\textbf{Organization.} \S 2 discusses related works. In \S 3 we give a formal definition of the session-based social recommendation problem. Our session-based social recommendation approach is described in \S 4. \S 5 presents the experimental results, followed by concluding remarks in \S 6.

%% file: related.tex
\section{Related Work}
We discuss three lines of research that are relevant to our work: 1) recommender systems that model the dynamic user behaviors, 2) social recommender systems that take social influence into consideration, and 3) recent progress of convolutional network developed for graph-structured data.

\subsection{Dynamic Recommendation}
Modeling user interests that change over time has already received some attention~\cite{xiong2010temporal,koren2010collaborative,charlin2015dynamic}.
Most of these models are based on (Gaussian) matrix factorization \cite{mnih2008probabilistic}.
For example, \citet{xiong2010temporal} learned temporal representations by factorizing the (user, item, time) tensor. \citet{koren2010collaborative} developed a similar model named timeSVD++. \citet{charlin2015dynamic} modeled similarly but using Poisson factorization~\cite{gopalan2015scalable}. 
However, these approaches assume that the interest of users changes slowly and smoothly over long-term horizons, typically on the order of months or years.
To effectively capture users' short-term interests, recent works introduce RNN to model their recent (ordered) behaviors. For example, \citet{hidasi2016session} first proposed Session-RNN to model user's interest within a session. \citet{li2017neural} further extended Session-RNN with attention mechanism to capture user's both local and global interests. \citet{wu2017recurrent} used two separate RNNs to update the representations of both users and items based on new observations. \citet{beutel2018latent} built an RNN-based recommender that can incorporate auxiliary context information. These models assume that items exhibit coherence within a period of time, and we use a similar approach to model session-based user interests.


\subsection{Social Recommendation}
Modeling the influence of friends on user interests has also received attention~\cite{massa2007trust,ma2008sorec,ma2011recommender,jamali2010matrix,jiang2012social}. Most proposed models are (also) based on Gaussian or Poisson matrix factorization. For example, \citet{ma2011recommender} studied social recommendations by regularizing latent user factors such that the factors of connected users are close by.
\citet{chaney2015probabilistic} weighted the contribution of friends on a user's recommendation using a learned ``trust factor''. 
\citet{zhao2014leveraging} proposed an approach to leverage social networks for active learning. 
\citet{xiao2017learning} framed the problem as transfer learning between the social domain and the recommendation domain. These approaches can model social influences assuming influences are uniform across friends and independent from the user's preferences. 
\citet{tang2012etrust} and \citet{tang2012mtrust} proposed multi-facet trust relations, which relies on additional side information (e.g., item category) to define facets. \citet{wang2016social} and \citet{wang2017learning} distinguished strong and weak ties among users for recommendation in social networks. However, they ignore the user's short-term behaviors and integrate context-independent social influences.
Our proposed approach models dynamic social influences by modeling the dynamic user interests, and context-dependent social influences.

\subsection{Graph Convolutional Networks}
Graph convolutional networks (GCNs) inherits convolutional neural networks (CNNs).
CNNs have achieved great success in computer vision and several other applications. CNNs are mainly developed for data with 2-D grid structures such as images~\cite{krizhevsky2012imagenet}. 
Recent works focus on modeling more general graph-structure data using CNNs~\cite{bruna2013spectral,henaff2015deep,defferrard2016convolutional,kipf2016semi}. Specifically, \citet{kipf2016semi} proposed graph-convolutional networks (GCNs) for semi-supervised graph classification. The model learns node representations by leveraging both the node attributes and the graph structure. It is composed of multiple \emph{graph-convolutional layers}, each of which updates node representations using a combination of the current node's representation and that of its neighbors. Through this process, the dependency between nodes is captured. However, in the original formulation, all neighbors are given the static ``weight'' when updating the node representations. \citet{Velickovic2018graph} addressed this problem by proposing \emph{graph-attention networks}. They weighed the contribution of neighbors differently using an attention mechanism~\cite{bahdanau2015neural,xu2015show}.

We propose a dynamic-graph-attention network. Compared to previous work, we focus on a different application (modeling the context-dependent social influences for recommendations). Besides, we model a dynamic graph, where the features of nodes evolve over time, and the attention between nodes also changes along with the current context over time. 

%% file: definition.tex
\section{Problem Definition}

Recommender systems suggest relevant items to their users according to their historical behaviors. In classical recommendation models (e.g., matrix factorization \cite{mnih2008probabilistic}), the order in which a user consumes items is ignored. However, in online communities, user-preferences change rapidly, and the order of user preference behaviors must be considered so as to model users' dynamic interests. In practice, since users' entire history record can be extremely long (e.g., certain online communities have existed for years) and users' interests switch quickly, a common approach is to segment user preference behaviors into different sessions (e.g., using timestamps and consider each user's behavior within a week as a session) and provide recommendations at session level~\cite{hidasi2016session}. We define this problem as follows:

\textup{DEFINITION 1.} \textbf{(Session-based Recommendation)} Let $U$ denote the set of users and $I$ be the set of items. 
Each user $u$ is associated with a set of sessions by the time step $T$, $I^u_T=\{ \vec{S}_1^u, \vec{S}_2^u,\ldots, \vec{S}_T^u\}$, 
where $\vec{S}_t^u$
is the $t_{th}$ session of user $u$. Within each session, $\vec{S}_t^u$ consists of a sequence of user behaviors $\{i_{t,1}^{u}, i_{t,2}^{u},\ldots, i_{t,N_{u,t}}^{u}\}$, where $i_{t,p}^{u}$ is the $p_{th}$ item consumed by user $u$ in $t_{th}$ session, and $N_{u,t}$ is the amount of items in the session. For each user $u$, given a new session $\vec{S}_{T+1}^u=\{i_{T+1,1}^{u},\ldots, i_{T+1,n}^{u}\}$, the goal of \emph{session-based recommendation} is to recommend a set of items from $I$ that the user is likely to be interested in during the next step $n+1$, i.e., $i_{T+1,n+1}^{u}$.

In online communities, users' interests are not only correlated to their historical behaviors, but also commonly influenced by their friends.
For example, if a friend watches a movie, I may also be interested in watching it. This is known as social influence~\cite{tang2009social}. Moreover, the influences from friends are \emph{context-dependent}. In other words, the influences from friends vary from one situation to another. For example, if a user wants to buy a laptop, she will be more likely referring to friends who are keen on high-tech devices; while she may be influenced by photographer friends when shopping a camera. Like as Figure 1, a user can be influenced by both her friends' short- and long-term preferences. 

To provide an effective recommendation to users in online communities, we propose to model both users' dynamic interests and context-dependent social influences. We define the resulting problem as follows:

\textup{DEFINITION 2.} \textbf{(Session-based Social Recommendation)} Let $U$ denote the set of users, $I$ be the set of items, and $G=(U, E)$ be the social network, where $E$ is the set of social links between users. Given a new session $\vec{S}_{T+1}^u=\{i_{T+1,1}^{u},\ldots, i_{T+1,n}^{u}\}$ of user $u$, the goal of \emph{session-based social recommendation} is to recommend a set of items from $I$ that $u$ is likely to be interested in during the next time step $n+1$ by utilizing information from both her dynamic interests (i.e., information from $\cup_{t=1}^{T+1}\vec{S}_t^u$) and the social influences (i.e., information from $\cup_{k=1}^{N(u)}\cup_{t=1}^{T}\vec{S}_t^k$, where $N(u)$ is the set of $u$'s friends).

%% file: model.tex
\begin{figure*}
\centering
\includegraphics[width=0.9\textwidth]{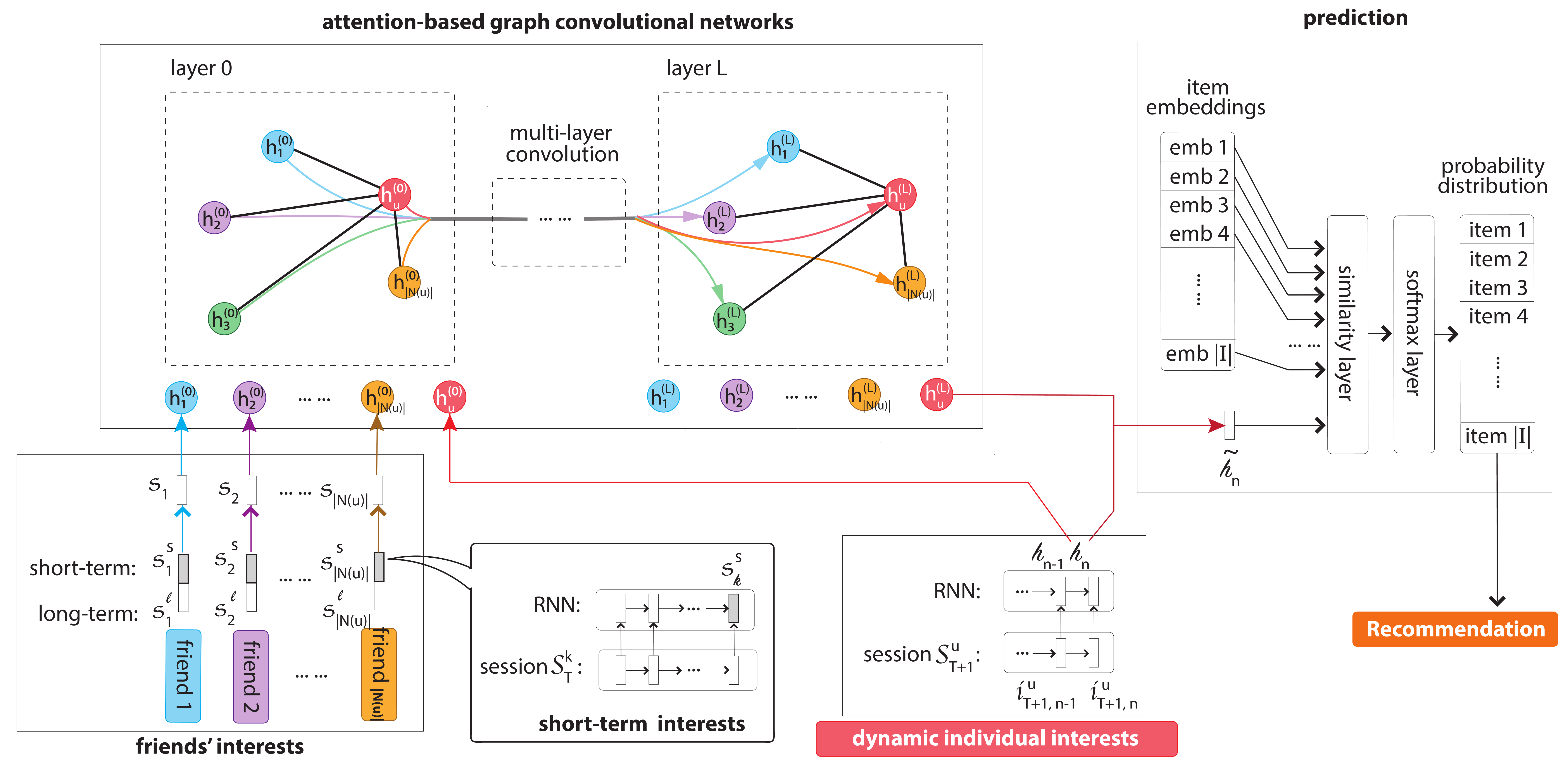}
\vspace{-8pt}
\caption{A schematic view of our proposed model for dynamic social recommendation. 
}
\label{fig:overview_model}
\end{figure*}

\section{Dynamic Social Recommender Systems}
As is discussed previously, users are not only guided by their current preferences but also by their friends' preferences. 
We propose a novel dynamic graph attention model \gls{DGRec} which models both types of preferences. 

\gls{DGRec} is composed of four modules (Figure~\ref{fig:overview_model}).
First (\S \ref{sec:dynamic_individual_interests}), a recurrent neural network (RNN)~\cite{elman1990finding} models the sequence of items consumed in the (target) user's current session. 
Her friends' interests are modeled using a combination of their short- and long-term preferences (\S \ref{sec:rep_of_friends}).
The short-term preferences, or items in their most recent session, are also encoded using an RNN. Friends' long-term preferences are encoded with a learned individual embedding.
The model then combines the representation of the current user with the representations of her friends using a graph-attention network (\S \ref{sec:context_dependent_social_influences}). This is a key part of our model and contribution: our proposed mechanism learns to weigh the influence of each friend based on the user's current interests.
At the final step (\S \ref{sec:recommendation}), the model produces recommendations by combining a user's current preferences with her (context-dependent) social influences.

\subsection{Dynamic Individual Interests} \label{sec:dynamic_individual_interests}
To capture a user's rapidly-changing interests, we use RNN to model the actions (e.g., clicks) of the (target) user in the current session.
RNN is standard for modeling sequences and has recently been used for modeling user (sequential) preference data~\cite{hidasi2016session}. The RNN infers the representation of a user's 
session $\vec{S}_{T+1}^u=\{i_{T+1,1}^{u},\ldots, i_{T+1,n}^{u}\}$, token by token by recursively combining the representation of all previous tokens with the latest token, i.e., 
\begin{equation}\label{eqt:rnn}
h_n = f(i_{T+1,n}^{u}, h_{n-1}),
\end{equation}
where $h_n$ represents a user's interests and $f(\cdot,\cdot)$ is a non-linear function combining both sources of information. In practice, the long short-term memory (LSTM)~\cite{hochreiter1997long} unit is often used as the combination function $f(\cdot,\cdot)$:
\begin{equation}
\label{eqt:lstm}
\begin{aligned}
x_n &= \sigma(\textbf{W}_x[h_{n-1}, i_{T+1,n}^{u}] + b_x)\\
f_n &= \sigma(\textbf{W}_f[h_{n-1}, i_{T+1,n}^{u}] + b_f)\\
o_n &= \sigma(\textbf{W}_o[h_{n-1}, i_{T+1,n}^{u}] + b_o)\\
\tilde{c}_n &= \tanh(\textbf{W}_c[h_{n-1}, i_{T+1,n}^{u}] + b_c)\\
c_n &= f_n \odot c_{n-1} + x_n \odot \tilde{c}_n\\
h_n &= o_n \odot \tanh(c_n),\\
\end{aligned}
\end{equation}
where $\sigma$ is the sigmoid function: $\sigma(x) =(1+\exp(-x))^{-1}$. 

\subsection{Representing Friends' Interests}
\label{sec:rep_of_friends}
We consider both friends' short- and long-term interests. 
Short-term interests are modeled using the sequence of recently-consumed items (e.g., a friend's latest online session).
Long-term interests represent a friend's average interest and are modeled using individual embedding.

\textbf{Short-term preference}: For a target user's current session $\vec{S}_{T+1}^u$, her friends' short-term interests are represented 
using their sessions right before session $T+1$ (our model generalizes beyond single session but this is effective empirically). Each friend $k$'s actions $\vec{S}_{T}^k=\{i_{T,1}^{k}, i_{T,2}^{k},\ldots, i_{T,N_{k,T}}^{k}\}$ are modeled using an RNN. In fact, here we reuse the RNN for modeling the target user's session (\S~\ref{sec:dynamic_individual_interests}). In other words, both RNNs share the same weights.
We represent friend $k$'s short-term preference $s_k^s$ by the final output of the RNN:
\begin{equation}
\begin{aligned}
s_k^s &= r_{N_{k,T}} = f(i_{T, N_{k,T}}^{k}, r_{N_{k,T}-1}).
\end{aligned}
\end{equation}

\textbf{Long-term preference}: Friends' long-term preferences reflect their average interests.
Since long-term preferences are not time-sensitive, we use a single vector to represent them. Formally,
\begin{equation}
s_k^l = \mathbf{W}_u[k,:],
\end{equation}
where friend $k$'s long-term preference $s_k^l$ is the $k_{th}$ row of the user embedding matrix $\textbf{W}_u$.

Finally, we concatenate friends' short- and long-term preferences using a non-linear transformation:
\begin{equation}\label{eqa:long_short}
s_k = ReLU(\mathbf{W}_1[s_k^{s};s_k^{l}]),
\end{equation}
where $ReLU(x)=max(0,x)$ is a non-linear activation function and $\textbf{W}_1$ is the transformation matrix.

\subsection{Context-dependent Social Influences}\label{sec:context_dependent_social_influences}
We described how we obtain representations of target user (\S~\ref{sec:dynamic_individual_interests}) and her friends (\S~\ref{sec:rep_of_friends}). We now combine both into a single representation that we then use downstream (\S \ref{sec:recommendation}). The combined representation is a mixture of the target user's interest and her friends' interest. 

We obtain this combined representation using a novel graph-attention network. 
First, we encode the friendship network in a graph where nodes correspond to users (i.e., target users and their friends) and edges denote friendship. In addition, each node uses its corresponding user's representation (\S \ref{sec:dynamic_individual_interests} \& \S\ref{sec:rep_of_friends}) as (dynamic) features. Second, these features are propagated along the edges using a message-passing algorithm~\cite{gilmer2017neural}. The main novelty of our approach lies in using an attention mechanism to weigh the features traveling along each edge. A weight corresponds to the level of a friend's influence. After a fixed number of iterations of message passing, the resulting features at the target user's node are the combined representation. 

Below we detail how we design the node features as well as the accompanying graph-attention mechanism.

\subsubsection{Dynamic feature graph}
For each user, we build a graph where nodes correspond to that user and her friends. For target user $u$ with $|N(u)|$ friends, the graph has $|N(u)|+1$ nodes. 
User $u$'s initial representation $h_n$ is used as node $u$'s features $h_u^{(0)}$ (the features are updated whenever $u$ consumes a new item in $\vec{S}_{T+1}^u$).
For a friend $k$, the corresponding node feature is set to $s_k$ and remains unchanged for the duration of time step $T+1$. Formally, the node features are $h_u^{(0)}=h_n$ and \{$h_{k}^{(0)} = s_k, k\in{N(u)}$\}.

\subsubsection{Graph-Attention Network} With the node features defined as above, we then pass messages (features) to combine friends' and the target user's interests. This procedure is formalized as inference in a graph convolutional network~\cite{kipf2016semi}. 

\citet{kipf2016semi} introduce graph convolutional networks for semi-supervised node representation learning. In these networks, the convolutional layers ``pass'' the information between nodes. The number of layers $L$ of the networks corresponds to the number of iterations of message passing.\footnote{We propagate information on a graph that also contains higher-order relationships (e.g., friends of friends of friends) in practice. In the $l^{th}$ layer of the network, the target user then receives information from users that are $l$ degrees away.}
However, all neighbors are treated equally.
Instead, we propose a novel dynamic graph attention network to model context-dependent social influences.

\begin{figure}
\centering
\includegraphics[width=0.95\linewidth]{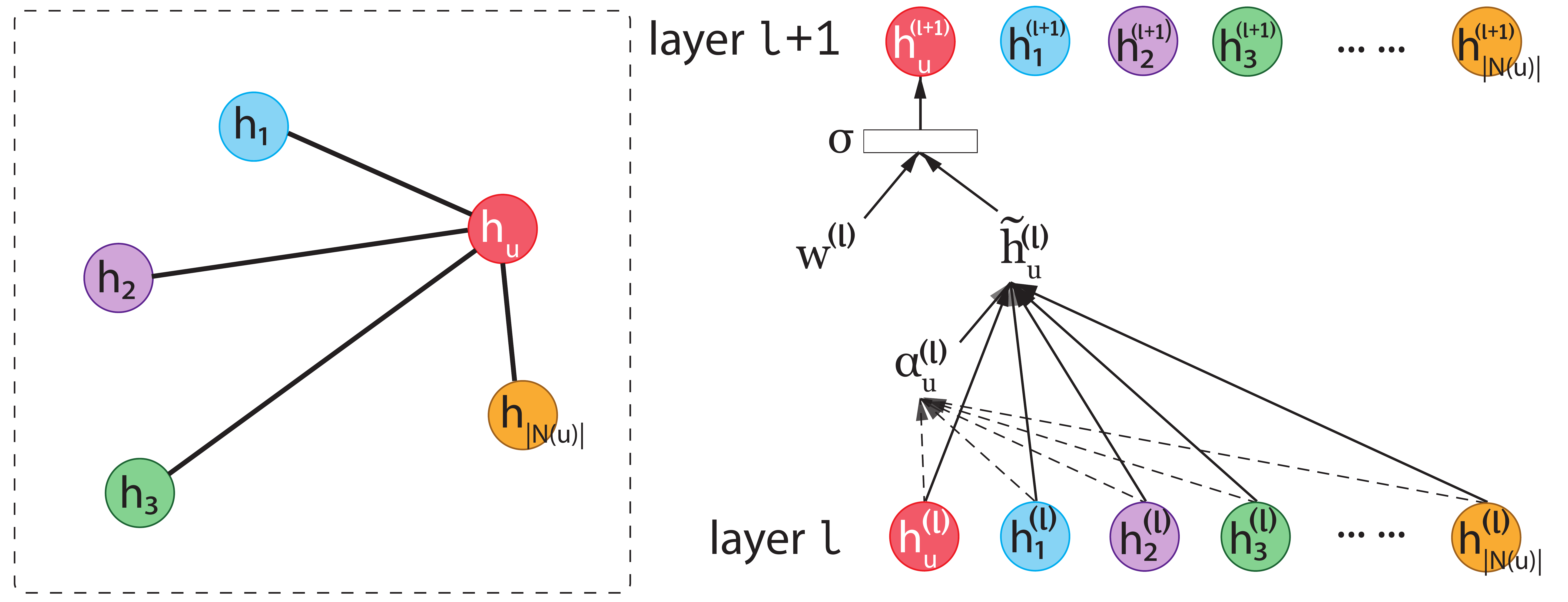}
\caption{The graphical model of the single convolutional layer using attention mechanism, where the output conditioned on current interest is interpreted as context-dependent social influences.}
\label{fig:attention_module}
\end{figure}

The fixed symmetric normalized Laplacian
is widely used as a propagation strategy in existing graph convolutional networks~\cite{defferrard2016convolutional,kipf2016semi}. In order to distinguish the influence of each friend, we must break the static propagation schema first.
We propose to use an attention mechanism to guide the influence propagation. The process is illustrated in Figure~\ref{fig:attention_module}. We first calculate the similarity between the target user's node representation $h_u^{(l)}$ and all of its neighbors' representations $h_k^{(l)}$:
\begin{equation}\label{eqt:alignment}
\alpha_{uk}^{(l)} = \frac{exp(f(h_u^{(l)},h_k^{(l)}))}{\sum_{j\in{N(u)\cup\{u\}}}exp(f(h_u^{(l)},h_j^{(l)}))},
\end{equation}
where $h_u^{(l)}$ is the representation of node/user $u$ at layer $l$, and $f(h_u^{(l)},h_k^{(l)}) = {h_u^{(l)}}^\top h_k^{(l)}$ is the similarity function between two elements. Intuitively, $\alpha_{uk}^{(l)}$ is the \emph{level of influence} or weight of friend $k$
on user $u$ (conditioned on the current context $h_u^{(l)}$). Note that we also include a self-connection edge to preserve a user's revealed interests. 
$\alpha_{u:}^{(l)}$ then provide the weights to combine the features:
\begin{equation}
\tilde{h}_u^{(l)} = \sum_{k\in{N(u)\cup\{u\}}}\alpha_{uk}^{(l)}h_k^{(l)},
\end{equation}
where $\tilde{h}_u^{(l)}$ is a mixture of user $u$'s friends' interests at layer $l$, followed by a non-linear transformation:
$h_u^{(l+1)} = ReLU(\textbf{W}^{(l)}\tilde{h}_u^{(l)}).$ $\textbf{W}^{(l)}$ is the shared and learnable weight matrix at layer $l$. We obtain the final representation of each node by stacking this attention layer $L$ times.\footnote{We also tested our model with two popular context-independent propagation strategies that do not use an attention mechanism: a) averaging friends' interests and; b) element-wise max-pooling over their interests---similar to techniques for aggregating word-level embeddings \cite{weston2014tagspace}. Mean aggregation outperforms the latter, but both are inferior to our proposed attention model.} The combined (social-influenced) representation is denoted by $h_u^{(L)}$.

\subsection{Recommendation}\label{sec:recommendation}
Since a user's interest depends on both her recent behaviors and social influences, her final representation is obtained by combining them using a fully-connected layer:
\begin{equation}\label{eqt:attnlayer}
\hat{h}_n = \textbf{W}_2[h_n; h_u^{(L)}],
\end{equation}
where $\textbf{W}_2$ is a linear transformation matrix, and $\hat{h}_n$ is the final representation of the user $u$'s current interest. 

We then obtain the probability that the next item will be $y$ using a softmax function:
\begin{equation}\label{eqt:softmax2}
p(y | i_{T+1,1}^{u},\ldots, i_{T+1,n}^{u}; \{\vec{S}_{T}^k, k\in{N(u)}\}) = \frac{\exp({\hat{h}_n}^\top z_y)}{\sum_{j=1}^{|I|}\exp({\hat{h}_n}^\top z_j)}, 
\end{equation}
where $N(u)$ are user $u$'s set of friends according to the social network $G$, $z_y$ is the embedding of item $y$, and $|I|$ the total number of items.

\subsection{Training}
We train the model by maximizing the log-likelihood of the observed items in all user sessions:
\begin{equation}
\label{eqn::obj}
   \sum_{u\in U} \sum_{t=2}^{T} \sum_{n=1}^{N_{u, t}-1} \log p(i_{t,n+1}^{u}|i_{t,1}^{u},\ldots, i_{t,n}^{u}; \{\vec{S}_{t-1}^k, k\in{N(u)}\}).
\end{equation}
This function is optimized using gradient descent. 

%% file: experiment.tex
\section{Experiments}
Studying the effectiveness of our \gls{DGRec} using real-world data sets, we highlight the following results: 
\begin{itemize}
\item \gls{DGRec} significantly outperforms all seven methods that it is compared to under all experimental settings.
\item Ablation studies demonstrate the usefulness of the different components of \gls{DGRec}.
\item Exploring the fitted models shows that attention contextually weighs the influences of friends.
\end{itemize}
\subsection{Experimental Setup}
\subsubsection{Data Sets}
We study all models using data collected from three well-known online communities. Descriptive statistics for all data sets are in Table~\ref{tab::stat}.

\textit{Douban.}\footnote{http://www.douban.com} A popular site on which users can review movies, music, and books they consume. We crawled the data using the identities of the users in the movie community, obtaining every movie they reviewed along with associated timestamps. We also crawled the users' social networks. We construct our data set by using each review as an evidence that a user consumed an item. Users tend to be highly active on Douban so we segment users' behaviors (movie consumption) into week-long sessions.

\textit{Delicious.}\footnote{Data set available from {\url{https://grouplens.org/datasets/hetrec-2011/}}} An online bookmarking system where users can store, share, and discover web bookmarks and assign them a variety of semantic tags. The task we consider is personalized tag recommendations for bookmarks. Each session is a sequence of tags a user has assigned to a bookmark (tagging actions are timestamped). This differs from the ordinary definition of sessions as a sequence of consumptions over a short horizon.

\textit{Yelp.}\footnote{Data set available from {\url{https://www.yelp.com/dataset}}}
An online review system where users review local businesses (e.g., restaurants and shops). Similar as for Douban, we treat each review as an observation. Based on the empirical frequency of the reviews, we segment the data into month-long sessions.

We also tried different segmentation strategies. Preliminary results showed that our method consistently outperformed Session-RNN and NARM for other session lengths. We leave a systematic study for optimizing session segmentation as our future work.

\subsubsection{Train/valid/test splits} We reserve the sessions of the last $d$ days for testing and filter out items that did not appear in the training set. Due to the different sparseness of the three data sets, we choose $d = 180, 50$ and $25$ for \textit{Douban}, \textit{Yelp} and \textit{Delicious} data sets respectively. We randomly and equally split the held out sessions into validation and test sets.

\begin{table}
\centering
\begin{tabular}{lccc} 
\toprule 
 & Douban & Delicious & Yelp \\
\midrule 
\#\ Users & 32,314 & 1,650 & 141,804 \\
\#\ Items & 14,109 & 4,282 & 17,625 \\
\#\ Events & 3,493,821 & 296,705 & 1,200,503 \\
\#\ Social links & 331,315 & 15,328 & 6,818,026 \\
Start Date & 01/12/2008 & 08/12/2009 & 01/01/2009\\
End Date & 07/22/2016 & 07/01/2016 & 10/15/2010 \\
\midrule 
Avg. friends/user & 10.25 & 9.00 & 48.08 \\
Avg. events/user & 108.12 & 179.82 & 8.47 \\
Avg. session length & 4.38 & 4.30 & 3.63 \\
\bottomrule 
\end{tabular}
\caption{Descriptive statistics of our three data sets.}
\vspace{-15pt}
\label{tab::stat}
\end{table}

\begin{table*}
\centering
\begin{tabular}{llcccccc}
\toprule 
\multirow{2}{*}{Model Class} & \multirow{2}{*}{Model} & \multicolumn{2}{c}{Douban} & \multicolumn{2}{c}{Delicious} & \multicolumn{2}{c}{Yelp} \\
& & Recall@20 & NDCG & Recall@20 & NDCG & Recall@20 & NDCG \\
\midrule 
\multirow{2}{*}{Classical} & ItemKNN~\cite{linden2003amazon} & 0.1431 & 0.1635 & 0.2729 & 0.2241 & 0.0441 & 0.0989 \\
 & BPR-MF~\cite{rendle2009bpr} & 0.0163 & 0.1110 & 0.2775 & 0.2293 & 0.0365 & 0.1190 \\
 \midrule 
\multirow{3}{*}{Social} & SoReg~\cite{ma2011recommender} & 0.0177 & 0.1113 &0.2703 & 0.2271 & 0.0398 & 0.1218 \\ 
& SBPR~\cite{zhao2014leveraging} & 0.0171 & 0.1059 & 0.2948 & 0.2391 & 0.0417 & 0.1207\\
& TranSIV~\cite{xiao2017learning} & 0.0173 & 0.1102 & 0.2588 & 0.2158 & 0.0420 & 0.1187 \\
\midrule 
\multirow{2}{*}{Temporal} & RNN-Session~\cite{hidasi2016session} & 0.1643 & 0.1854 & 0.3445 & 0.2581 & 0.0756 & 0.1378 \\
& NARM~\cite{li2017neural} &0.1755 &0.1872 &0.3776 &0.2768 &0.0765 & 0.1380 \\
\midrule 
Social + Temporal (Ours) & DGRec & \textbf{0.1861} & \textbf{0.1950} & \textbf{0.4066} & \textbf{0.2944} & \textbf{0.0842} & \textbf{0.1427} \\
\bottomrule 
\end{tabular}
\caption{Quantitative Results of Different Algorithms. We highlight that \gls{DGRec} outperforms all other baselines across all three data sets and both metrics. Further analysis is provided in \S \ref{sec:quantitative_results}.}
\vspace{-10pt}
\label{table:results}
\end{table*}

\subsubsection{Competing Models}\label{sec:models}

We compare \gls{DGRec} to three classes of recommenders: (A) classical methods that utilize neither social nor temporal factors; (B) social recommenders, which take context-independent social influences into consideration; and (C) session-based recommendation methods, which model user interests in sessions. (Below, we indicate a model's class next to its name.)

\begin{itemize}[leftmargin=0.2in]
\item ItemKNN~\cite{linden2003amazon} (A): inspired by the classic KNN model, it looks for items that are similar to items liked by a user in the past.
\item BPR-MF~\cite{rendle2009bpr} (A): matrix factorization (MF) technique trained using a ranking objective as opposed to a regression objective.
\item SoReg~\cite{ma2011recommender} (B): uses the social network to regularize the latent user factors of matrix factorization.
\item SBPR~\cite{zhao2014leveraging} (B): an approach for social recommendations based on BPR-MF. The social network is used to provide additional training samples for matrix factorization.
\item TranSIV~\cite{xiao2017learning} (B): uses shared latent factors to transfer the learned information from the social domain to the recommendation domain.
\item RNN-Session~\cite{hidasi2016session} (C): recent state-of-the-art approach that uses recurrent neural networks for session-based recommendations.
\item NARM~\cite{li2017neural} (C): a hybrid model of both session-level preferences and the user's ``main purpose'', where the main purpose is obtained via attending on previous behaviors within the session.
\end{itemize}

\subsubsection{Evaluation Metrics}
We evaluate all models with two widely used ranking-based metrics: Recall@K and Normalized Discounted Cumulative Gain (NDCG).

\textit{Recall@K} measures the proportion of the top-K recommended items that are in the evaluation set. We use $K=20$.

\textit{NDCG} is a standard ranking metric. In the context of session-based recommendation, it is formulated as: $\text{NDCG}=\frac{1}{\log_{2}(1+\text{rank}_{pos})}$, where $\text{rank}_{pos}$ denotes the rank of a positive item. We report the average value of NDCG over all the testing examples.

\subsubsection{Hyper-parameter Settings}
For RNN-Session, NARM and our models, we use a batch size of 200. We use Adam~\cite{kingma2014adam} for optimization due to its effectiveness with $\beta_1=0.9$, $\beta_2=0.999$ and $\epsilon=1e^{-8}$ as suggested in TensorFlow~\cite{tensorflow2015-whitepaper}. The initial learning rate is empirically set to 0.002 and decayed at the rate of 0.98 every 400 steps. For all models, the dimensions of the user (when needed) and item representations are fixed to 100 following~\citet{hidasi2016session}. We cross-validated the number of hidden units of the LSTMs and the performance plateaued around 100 hidden units.
The neighborhood sample sizes are empirically set to 10 and 15 in the first and second convolutional layers, respectively. We tried to use more friends in each layer but observed no significant improvement. 
In our models, dropout~\cite{srivastava2014dropout} with rate $0.2$ is used to avoid overfitting.

\subsubsection{Implementation Details} We implement our model using TensorFlow~\cite{tensorflow2015-whitepaper}. Training graph attention networks on our data with mini-batch gradient descent is not trivial since node degrees have a large range. We found the neighbor sampling technique proposed in \cite{hamilton2017inductive} pretty effective. 
Further, to reasonably reduce the computational cost of training \gls{DGRec}, we represent friends' short-term interests using only their most recent sessions.

\subsection{Quantitative Results}\label{sec:quantitative_results}
The performance of different algorithms is summarized in Table~\ref{table:results}. ItemKNN and BPR-MF
perform very similarly, except on \textit{Douban}. 
A particularity of Douban is that users typically only consume each item once (different from \text{Delicious} and \text{Yelp}). MF-based methods tend to recommend previously consumed items which explain BPR-MF's poor performance.
By modeling social influence, the performance of social recommenders improves compared to BPR-MF in most cases. However, the improvement is marginal because these three algorithms (B) only model context-independent social influence. By modeling dynamic user interests, RNN-Session significantly outperforms ItemKNN and BPR, which is consistent with the results in \citet{hidasi2016session}.
Further, NARM extends RNN-Session by explicitly modeling user's main purpose and becomes the strongest baseline. Our proposed model \gls{DGRec} achieves the best performance among all the algorithms by modeling both user's dynamic interests and context-dependent social influences. Besides, the improvement over RNN-Session and NARM is more significant compared to that of SoReg over BPR-MF, which shows the necessity of modeling context-dependent social influences.

\subsection{Variations of \gls{DGRec}}
To justify and gain further insights into the specifics of \gls{DGRec}'s architecture, we now study and compare variations of our model. 
\subsubsection{Self v.s. Social} \gls{DGRec} obtains users' final preferences as a combination of user's consumed items in the current session and context-dependent social influences (see Eq.\ \ref{eqt:attnlayer}). To tease apart the contribution of both sources of information, we compare \gls{DGRec} against two submodels:
a) (\gls{DGRec}$_\text{self}$) a model of the user's current session only (Eq.\ \ref{eqt:attnlayer} without social influence features $h_{u}^{(L)}$) and; b) (\gls{DGRec}$_\text{social}$) a model using context-dependent social influence features only (Eq.\ \ref{eqt:attnlayer} without individual features $h_n$). Note that when using individual features only, \gls{DGRec}$_\text{self}$ is identical to RNN-Session (hence the results are reproduced from Table~\ref{table:results}).
Table~\ref{table:feature} reports the performance of all three models on our data sets.
\gls{DGRec}$_\text{self}$ consistently outperforms \gls{DGRec}$_\text{social}$ across all three data sets, which means that overall users' individual interests have a higher impact on recommendation quality.
Compared to the full model \gls{DGRec}, the performance of both \gls{DGRec}$_\text{self}$ and \gls{DGRec}$_\text{social}$ significantly decreases. To achieve good recommendation performance in online communities, it is, therefore, crucial to model both a user's current interests as well as her (dynamic) social influences.


\begin{table}
\centering
\begin{tabularx}{1.0\linewidth}{l>{\centering\arraybackslash}l>{\centering\arraybackslash}X>{\centering\arraybackslash}X}
\toprule
Data Sets & Models & Recall@20 & NDCG \\
\midrule
\multirow{3}{*}{\begin{tabular}[c]{@{}l@{}}Douban \end{tabular}} 
 & \gls{DGRec}$_\text{self}$ & 0.1643 & 0.1854 \\
 & \gls{DGRec}$_\text{social}$ & 0.1185 & 0.1591 \\
 & \gls{DGRec} & 0.1861 & 0.1950 \\
\midrule
\multirow{3}{*}{\begin{tabular}[c]{@{}l@{}}Delicious \end{tabular}} 
 & \gls{DGRec}$_\text{self}$ & 0.3445 & 0.2581 \\
 & \gls{DGRec}$_\text{social}$ & 0.3306 & 0.2516 \\
 & \gls{DGRec} & 0.4066 & 0.2944 \\
\midrule
\multirow{3}{*}{\begin{tabular}[c]{@{}l@{}}Yelp \end{tabular}} 
 & \gls{DGRec}$_\text{self}$ & 0.0756 & 0.1378 \\
 & \gls{DGRec}$_\text{social}$ & 0.0690 & 0.1356 \\
 & \gls{DGRec} & 0.0842 & 0.1427 \\
\bottomrule
\end{tabularx}
\caption{Ablation study comparing the performance of the complete model (\gls{DGRec}) with two variations.
}
\label{table:feature}
\end{table}

\begin{figure}
\centering
\vspace{-5pt}
\begin{subfigure}[t]{0.48\linewidth}
\includegraphics[width=\linewidth]{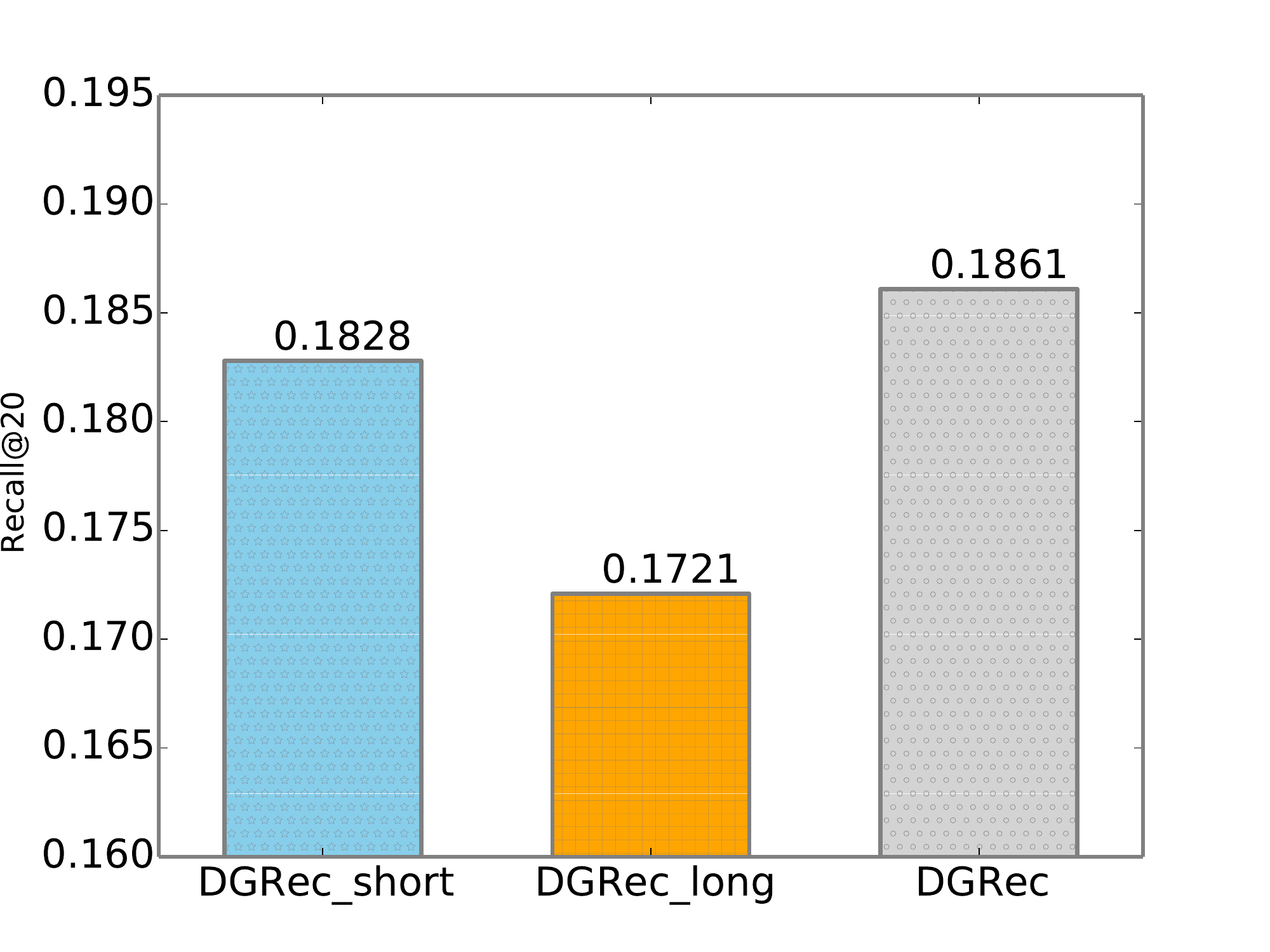}
\caption{Douban} \label{fig:locala}
\end{subfigure}
\hspace*{\fill} 
\begin{subfigure}[t]{0.48\linewidth}
\includegraphics[width=\linewidth]{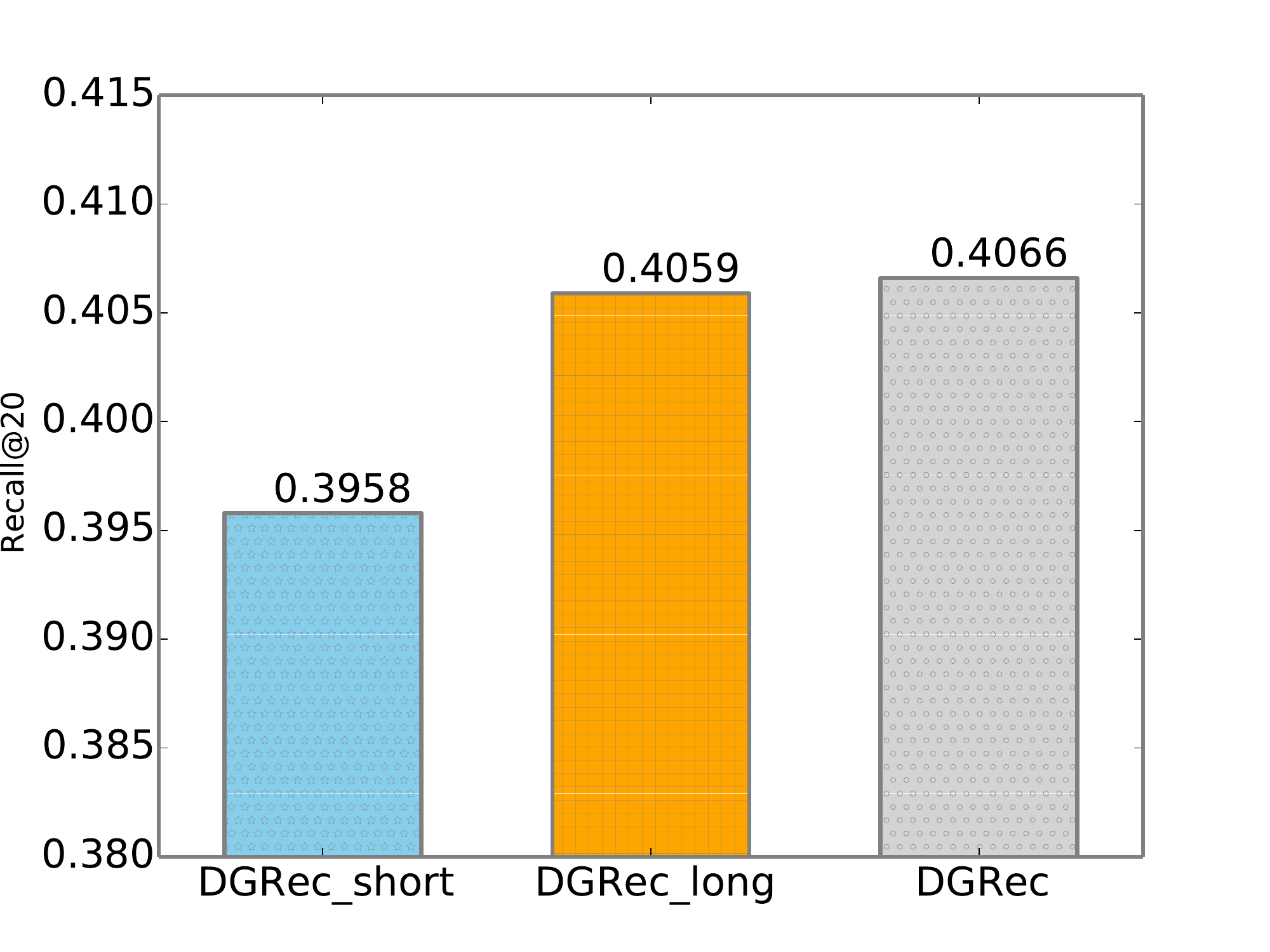}
\caption{Delicious} \label{fig:localc}
\end{subfigure}
\hspace*{\fill} 
\vspace{-10pt}
\caption{Performance w.r.t. friends' short-term and long-term preferences on different data sets. The result of \textit{Yelp} data set is similar with \textit{Douban} hence omitted.
}
\vspace{-10pt}
\label{fig:local_global}
\end{figure}

\begin{table}
\centering
\begin{tabularx}{1.0\linewidth}{l>{\centering\arraybackslash}X>{\centering\arraybackslash}X>{\centering\arraybackslash}X}
\toprule
Data Sets & Conv. Layers & Recall@20 & NDCG \\
\midrule
\multirow{3}{*}{\begin{tabular}[c]{@{}l@{}}Douban \end{tabular}} 
 & 1 & 0.1726 & 0.1886 \\
 & 2 & 0.1861 & 0.1950 \\
 & 3 & 0.1793 & 0.1894 \\
\midrule
\multirow{3}{*}{\begin{tabular}[c]{@{}l@{}}Delicious \end{tabular}} 
 & 1 & 0.4017 & 0.2883 \\
 & 2 & 0.4066 & 0.2944 \\
 & 3 & 0.4037 & 0.2932 \\
 \midrule
\multirow{3}{*}{\begin{tabular}[c]{@{}l@{}}Yelp \end{tabular}} 
 & 1 & 0.0760 & 0.1387 \\
 & 2 & 0.0842 & 0.1427 \\
 & 3 & 0.0846 & 0.1423 \\
\bottomrule
\end{tabularx}
\caption{Performance of our model w.r.t. different numbers of convolution layers.
}
\vspace{-20pt}
\label{table:conv_layer}
\end{table}

\subsubsection{Short-term v.s. Long-term}
\gls{DGRec} provides a mechanism for encoding friends' short- as well as long-term interests (see \S~\ref{sec:rep_of_friends}).
We study the impact of each on the model's performance. Similar to above, we compare using either short- or long-term interests to the results of using both.
Figure~\ref{fig:local_global} reports that for \textit{Douban}, the predictive capability of friends' short-term interests outperforms that of friends' long-term interest drastically, and shows comparable performance in regard to the full model. It is reasonable, considering that the interests of users in online communities (e.g., Douban) change frequently, and exploiting users' short-term interests should be able to predict user behaviors more quickly. Interestingly, on the data set \textit{Delicious}, different results are observed. Using long-term interests yield more accurate predictions than doing short-term. This is not surprising since, on \textit{Delicious} website, users tend to have static interests.

\subsubsection{Number of Convolutional Layers}
\gls{DGRec} aggregates friends' interests using a multi-layer graph convolutional network. More convolutional layers will yield influences from higher-order friends. In our study so far we have used two-layer graph convolutional networks. 
To validate this choice we compare the performance to one- and three-layer networks but maintain the number of selected friends to 10 and 5 in the first and third layer, respectively.
Table \ref{table:conv_layer} shows a significant decline in performance when using a single layer. This implies that the interests of friends' friends 
(obtained by 2 layers) 
is important for recommendations. 

Next, we test our model using three convolutional layers to explore the influences of even higher-order friends. The influence of the third layer on the performance is small. There is a small improvement for \textit{Yelp} but a slightly larger drop in performance for both \textit{Douban} and \textit{Delicious}, which may be attributed to
model overfitting or noises introduced by higher-order friends. This confirms that two convolutional layers are enough for our data sets.

\begin{figure}
\centering
\vspace{-8pt}
\includegraphics[width=0.8\linewidth]{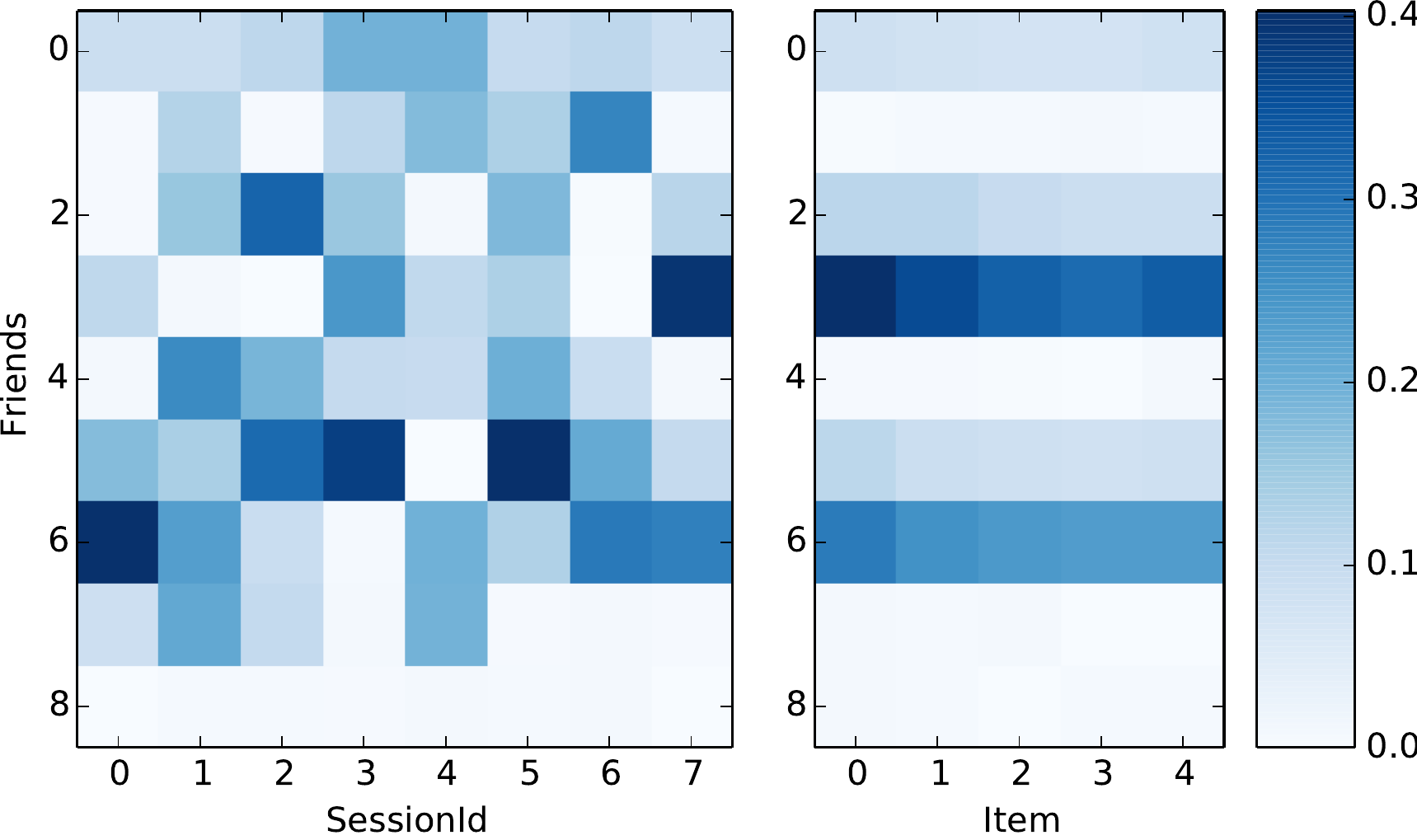}
\vspace{-8pt}
\caption{The heat map of the attention weights across different sessions (left) and within a session (right). For both plots, the y-axis represents friends of the target user. The x-axis represents (1) eight sessions of the target user on the left and (2) the item sequence within session \#7 on the right.}
\label{fig:attention}
\end{figure}

\begin{figure}
\centering
\vspace{-8pt}
\includegraphics[width=0.8\linewidth]{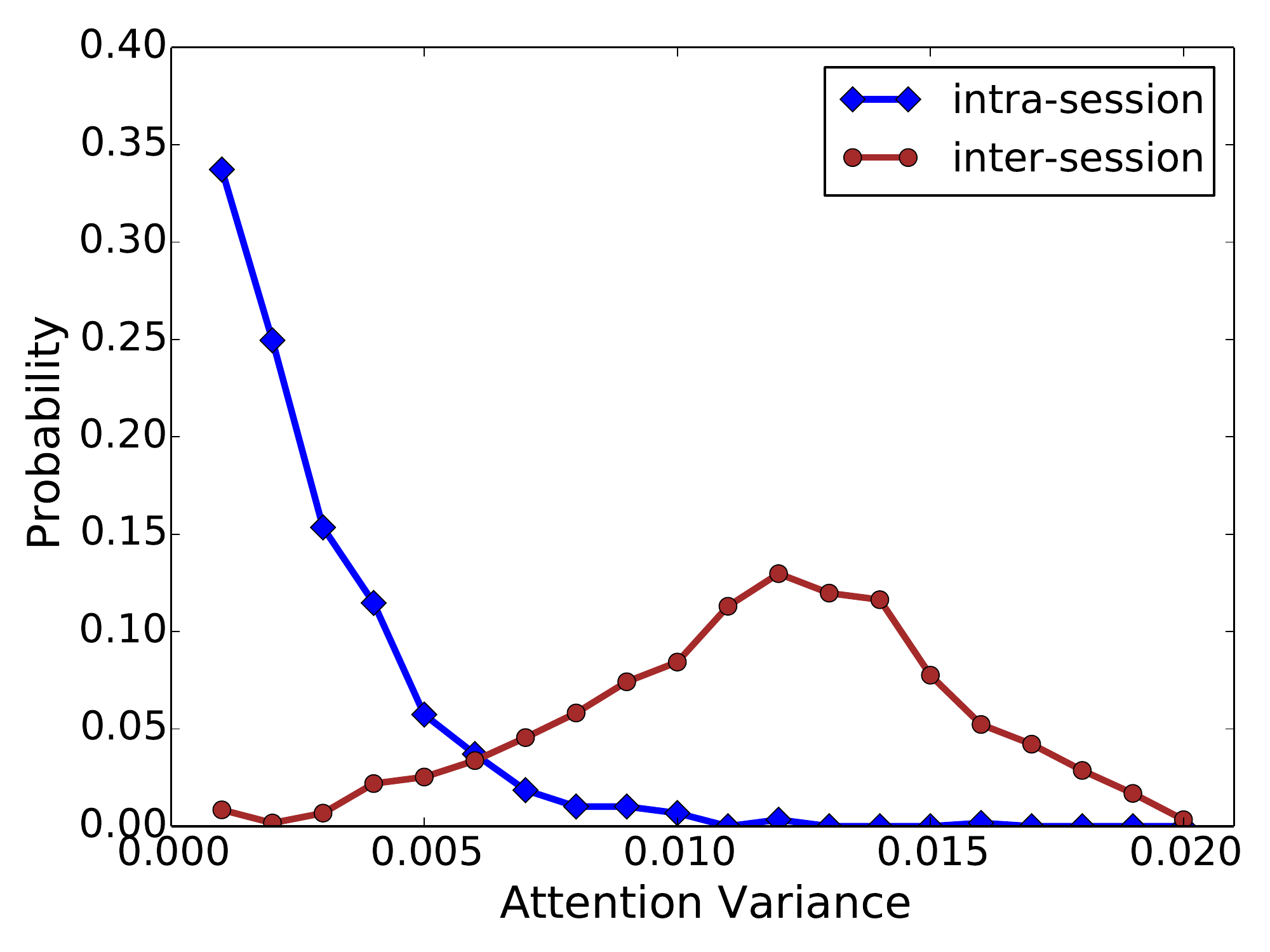}
\vspace{-8pt}
\caption{Attention variance distribution of \gls{DGRec} for both inter-session and intra-session. Variance values are discretized into 20 intervals.}
\vspace{-10pt}
\label{fig:attention_var}
\end{figure}

\subsection{Exploring Attention}
\gls{DGRec} uses an attention mechanism to weigh the contribution of different friends based on a user's current session. We hypothesized that while friends have varying interests, user session typically only explores a subset of these interests. As a consequence, for a target user, different subsets of her friends should be relied upon in different situations.
We now explore the results of the attention learned by our model.

First, we randomly select a \textit{Douban} user from those who have at least 5 test sessions as well as 5 friends
and plot her attention weights (Eq.\ \ref{eqt:alignment}) within and across session(s) in Figure~\ref{fig:attention}. For the inter-session level plot (left), we plot the average attention weight of a friend within a session.
For intra-session level plot (right), the user's attention weights within one session (i.e., SessionId=7) are presented. We make the following observations. First, the user allocates her attention to different friends across different sessions. This indicates that social influence is indeed conditioned on context (i.e., target user's current interests). Further, friend \#8 obtains little attention in all sessions, which means that social links do not necessarily lead to observed shared interest. Second, the distribution of attention is relatively stable within a single session. This confirms that the user's behaviors are coherent in a short period and suitable to be processed in a session manner.

As a second exploration of the behavior of the attention mechanism we take a macro approach and analyze the attention across all users (as opposed to a single user across friends). 
We use the attention levels inferred on the \textit{Douban} test set. 
Figure~\ref{fig:attention_var} reports the empirical distributions of the inter-session (brown) and intra-session (blue) attention variance (i.e., how much does the attention weights vary in each case). The intra-session variance is lower on average. This agrees with our assumption that users' interests tend to be focused within a short time so that the same set of friends are attended to for the duration of a session. 
On the contrary, a user is more likely to trust different friends in different sessions, which further validates modeling context-dependent social influences via attention-based graph convolutional networks.

%% file: conclusion.tex
\section{Conclusions}
We propose a model based on graph convolutional networks for session-based social recommendation in online communities. Our model first learns individual user representations by modeling the users' current interests. Each user's representation is then aggregated with her friends' representations using a graph convolutional networks with a novel attention mechanism. The combined representation along with the user's original representation is then used to form item recommendations. Experimental results on three real-world data sets demonstrate the superiority of our model compared to several state-of-the-art models. Next steps involve exploring user and item features indicative of preferences and further improving the performance of recommender systems for online communities.

%% file: ack.tex
\section{Acknowledgement}
This paper is partially supported by Beijing Municipal Commission of Science and Technology under Grant No. Z181100008918005 as well as the National Natural Science Foundation of China (NSFC Grant Nos.61772039, 61472006 and 91646202). We would like to thank Haoran Shi for collecting Douban data used in this paper.